\documentclass[preprint,aps]{revtex4}
\usepackage{graphicx}
\usepackage{dcolumn}
\usepackage{bm}

\begin{document}
\title{Degeneracy analysis for a super cell of a photonic crystal and its application to the creation of band gaps}
\author{Liang Wu$^{1}$ and Sailing He$^{1,2}$}
\address{$^{1}$
Center for Optical and Electromagnetic Research,\\
State Key Laboratory for Modern Optical Instrumentation, \\
Zhejiang University, email:wl@coer.zju.edu.cn\\
Yu-Quan, Hangzhou 310027, P. R. China\\
$^{2}$Division of Electromagnetic Theory, Alfven Laboratory,  Royal Institute of Technology, \\
S-100 44 Stockholm, Sweden
}
\date{\today}

\begin{abstract}
A method is introduced to
analyze the degeneracy properties of the band structure of a photonic crystal
making use of the super cells.
The band structure associated with a super cell of a photonic crystal has
degeneracies at the edge of the Brillouin zone if the photonic crystal has some kind
of point group symmetry. Both E-polarization and H-polarization cases have the same
degeneracies for a 2-dimensional (2D) photonic crystal. Two theorems are given and proved.
These degeneracies can be lifted to create photonic band gaps by
changing the transform matrix between the super cell and the smallest unit cell which is caused by
changing the translation group symmetry of the photonic crystal (the point group symmetry
of the photonic crystal may keep unchanged). The existence of the photonic
band gaps
for many known 2D photonic crystals is explained through the degeneracy analysis.
Some structures with large band gaps are also found under the guidance of this
degeneracy analysis.

\end{abstract}





\maketitle

\section{Introduction}
Photonic crystals,  which are periodic arrangement of dielectric or metallic
materials, have attracted a wide attention recently in both physics and engineering
communities in view of their unique ability to control light propagation \cite{yablonovitch,mold,book1,nature}.
Many potential applications of photonic crystals rely on their
photonic band gaps (PBGs).  It is thus of great interest to design
photonic crystals with an absolute band gap (as large as possible),
particularly, for a given dielectric material.

Two-dimensional (2D) photonic crystals have attracted special attention since they are easier to fabricate.
Many 2D photonic crystals with large absolute band gaps have been found  \cite{PR,pw1,Qiu}.
Based on the difference of the filling factors between the dielectric band and the air band
(related to the distribution of the displacement field), the rule of thumb sometimes can be used to explain
the band gaps particularly at  low frequencies  \cite{book1,jaos}.  Due to
the complication of the differential operators  in electrodynamics (different field components are coupled to each other
even if the permittivity  $\varepsilon({\bf r})$ is separable), it is difficult to obtain  analytical (even approximate) solutions  for
the distribution of the displacement field (particularly at high frequencies). Therefore,   many photonic crystals
with large absolute band gaps can not be explained or found by  the rule of thumb \cite{pw2,square,Ander}.

Degeneracy lifting  is another explanation for absolute band gaps and even a method to create band gaps
\cite{DC,Ani1,Ani2,Ander,Ander2,XZ}. The degeneracy can be lifted by e.g. using hexagonal photonic structures \cite{DC},
using anisotropic materials \cite{Ani1,Ani2}, breaking the space group symmetry \cite{Ander,Ander2} and
changing the dielectric distribution without breaking the space group
symmetry \cite{XZ}.
Both accidental and normal degeneracies
can exist in a photonic band structure (see e.g. \cite{XZ}, this is different as compared to an electronic system).
To investigate the degeneracy properties of 2D photonic crystals, the $E$-polarization and $H$-polarization
are usually considered separately as suggested by \cite{DC,FG}.
It is very complicated to  predict where the degeneracy appears and how to break the
degeneracy.

In some cases, we don't have to rely on such an analysis.
In the present paper, we introduce a method  to create degeneracies first and then
break them to create band gaps by studying the band structure associated with a super cell
(instead of the unit cell as considered in the literature mentioned before by others).
In the band structure associated with a super cell, we can analyze how degeneracies are formed
and how to break them to create band gaps.

The point group symmetry of a photonic crystal is defined with respect to the point
with the highest symmetry.
For example, the point group symmetry
does not changed by adding columns at the corners of the unit cell for the
2D photonic crystals considered in  \cite{Ander} (they belong to the same point group symmetry
$C_{4v}$).
We notice that the translation group symmetry does not change either.
Thus the space group symmetry of the photonic crystal does not change at
all although the smallest unit
cell must include two rods after the additional rods are added.
It may be hard to understand the degeneracy breaking for the $H$-polarization at point
${\bf M}$ of the second and third bands as shown in \cite{Ander}
without careful analysis of the electromagnetic field distribution.
However, if we study the band structure associated with a super cell, the lifting of the degeneracy
and the creation of PBGs of such photonic crystals can be understood
with some tricks even when the space group symmetry of the photonic crystal
does not change.
For the example mentioned above, the photonic crystal
with the additional columns
at the corners can be treated as the result of changing the translation group symmetry
(keeping the point group symmetry unchanged)
from another photonic crystal with the additional columns having the save size with
the original column (as shown in Fig. 3(a) with the square column case), which also belongs to
$C_{4v}$ point group symmetry ).
Another view is presented to understand the idea of additional columns.
Not surprisingly, many known photonic crystal structure such as a chessboard structure \cite{chessboard},
a square lattice of square rods \cite{square} and even a triangular air hole structure \cite{book1}
can be somehow understood from this point of view.
(cf. the numerical example associated with Figs. 3-6 below).
By using such a degeneracy analysis associated with a super cell,
some structures with large band gaps are also found
in the present paper.

\section{Theorems for degeneracies in the band structure associated with a super cell}

A unit cell we consider here refers to the smallest periodic region in a photonic crystal.
If the periodic region includes
more than one unit cells, e.g. 2 unit cells, it is called a super cell.
First  we want to study the relation between  the band structure associated with the super cell and the original band structure (associated with the unit cell).

In general, we consider a three-dimensional (3D) photonic crystal with primitive lattice vectors
${\bf a_1}$, ${\bf a_2}$  and ${\bf a_3}$. The associated primitive reciprocal vectors ${\bf b_1}$, ${\bf b_2}$
and ${\bf b_3}$ are determined by
\begin{eqnarray}
{\bf b}_i  = 2\pi \frac{\sum\limits_{j,k = 1}^3{\epsilon _{ijk}{\bf a}_j\times{\bf a}_k }}{{{\bf a}_1\cdot
({\bf a}_2\times {\bf a}_3 )}},
\end{eqnarray}
where $\epsilon_{ijk}$ is the 3D Levi-Civita completely antisymmetric symbol.
The complete set of the reciprocal lattice vectors is written as $\{{\bf G}|{\bf G}=l_1{\bf b_1}+
l_2{\bf b_2}+l_3{\bf b_3}\}$, where $(l_1, l_2, l_3)$ are integers. We denote the first Brillouin zone formed
by these reciprocal lattice vectors $\{{\bf G}\}$ as zone A.

The primitive lattice vectors for a super cell are the linear combinations (with integer coefficients) of the primitive lattice
vectors for the unit cell, i.e., ${\bf a'}_i={\sum\limits_{j = 1}^3N_{ij}{\bf a}_j}$, $i,j=1,2,3$, where
$N_{ij}$ are  integers.
The corresponding primitive reciprocal vectors for the super cell are determined by
${\bf b}'_i = 2\pi \frac{\sum\limits_{j,k = 1}^3{\epsilon _{ijk}{\bf a}'_j\times{\bf a}'_k }}
{{{\bf a}'_1\cdot({\bf a}'_2\times {\bf a}'_3)}}$.
The integers $N_{ij}$ form a $3\times 3$ transform matrix with a positive determinant $\det(N)\equiv M >0$.

Since
\begin{eqnarray}
{\bf b}_i\cdot{\bf a}'_j&=&2\pi\frac{\sum\limits_{m,n = 1}^3{\epsilon _{imn}{\bf a}_m\times {\bf a}_n }}
{{{\bf a}_1\cdot ({\bf a}_2\times {\bf a}_3)}}\cdot \sum\limits_{l=1}^3 N_{jl}{\bf a}_l \nonumber \\
\end{eqnarray}
it follows from ${\bf a'}_i\cdot{\bf b'}_j=2\pi \delta_{ij}$ that
\begin{eqnarray}
{\bf b}_i  =\sum\limits_{j = 1}^3{N_{ji} {\bf b}'_j} =\sum\limits_{j= 1}^3{N^T_{ij} {\bf b}'_j},
\end{eqnarray}
where the superscript $T$ denotes the matrix transposition.
The set of the reciprocal lattice vectors associated with the super cell is
$\{{\bf G'}|{\bf G'}=\sum\limits_{j= 1}^3n_j{\bf b'}_j\}$.
Since ${\bf G}=\sum\limits_{i = 1}^3n_i{\bf b}_i=\sum\limits_{i,j = 1}^3n_iN^T_{ij}{\bf b'}_j$, one sees that $\{{\bf G}\}$ is
a subset of $\{{\bf G'}\}$. Note that the elements of $\{{\bf G}\}$  and $\{{\bf G'}\}$ are the integer grid points (they do not fill any continuous space) formed by the corresponding reciprocal lattice vectors. We denote the first Brillouin zone formed
by the reciprocal vectors $\{{\bf G'}\}$ as zone B.

{\em Lemma.} There exist a subset $\{\bar{\bf G}\}$ of $\{{\bf G'}\}$, which satisfies:

(i) $\{\bar{\bf G}\}\subset\{{\bf G}'\}$ and $\{\bar{\bf G}\}\cap\{{\bf G}\}={\bf 0}$.

(ii) There are $M$ elements in the set $\{\bar{\bf G}\}$ ($M$ is the determinant
of the matrix $N$) and the difference of any two of them does not belong to $\{{\bf G}\}$, i.e.,
(${\bar{\bf G}}_1-{\bar{\bf G}}_2\notin\{{\bf G}\}$).

(iii) Any ${\bf G'}\in \{{\bf G'}\}$ can be expressed as
\begin{eqnarray}
{\bf G'}=\bar{\bf G}+{\bf G},
\end{eqnarray}
where ${\bf G}\in\{{\bf G}\}, {\bf \bar G}\in\{\bar{\bf G}\}$.

The proof and a way to find the set $\{\bar{\bf G}\}$ are given in the appendix.

If we define the addition of vectors as the multiplication in a group theory, we can take $\{{\bf G'}\}$ as a group
and $\{{\bf G}\}$ as a subgroup. Then the vector ${\bf 0}$ is the unit element of the group.
From the group theory, one knows that $\{{\bf G'}\}$ is the union of all the cosets of the set $\{{\bf G}\}$.
The subset $\{\bar{\bf G}\}$ is used to  give the cosets.

With these reciprocal vectors, each eigenstate of the
electromagnetic field component ${H_{\bf k}}$ (with the wave vector ${\bf k}$ in the first Brillouin zone) in the photonic crystal
can be expressed in terms of the following Bloch series \cite{pw2},
\begin{eqnarray}
H_{\bf k}({\bf r}) = e^{i{\bf k} \cdot {\bf r}} \sum\limits_{\bf G} {H_{\bf G} e^{i{\bf G} \cdot {\bf r}} } .
\end{eqnarray}
The field component $H_{\bf k}$ satisfies the following equation,
\begin{eqnarray}
\Theta H_{\bf k}  = \frac{{\omega_{\bf k} ^2 }}{{c^2 }}H_{\bf k},
\end{eqnarray}
where the operator $\Theta$ can be easily derived from Maxwell's equations, and $c$ is the speed of light.

For any wave vector ${\bf k}$ in the ${\bf k}$ space of  a photonic crystal, one can find a wave vector in the first Brillouin zone which has the same eigenstate.
The difference of the two wave vectors should be a reciprocal vector.
Therefore, for any wave vector ${\bf k}$,  there exist a ${\bf G}\in\{{\bf G}\}$ so that
\begin{eqnarray}
{\bf k}_1={\bf k}-{\bf G}
\end{eqnarray}
 is in zone A (associated with the unit cell)
and a ${\bf G'}\in\{{\bf G}'\}$ so that
\begin{eqnarray}
{\bf k}_2={\bf k}-{\bf G'}
\end{eqnarray}
is in zone B (associated to the super cell).
We call ${\bf k}_1$ (in zone A) the counterpoint of ${\bf k}_2$ (in zone B) for the same photonic crystal. They denote the same eigenstate in the reciprocal vector spaces associated with the unit cell and the super cell, respectively.

For a fixed ${\bf k}_2 \in$ B, we define the set $\{{\bf K}_1|{\bf K}_1={\bf k}_2+\bar{\bf G}$, for all $\bar{\bf G}\in \{\bar{\bf G}\}\}_{{\bf k}_2}$. Clearly, there are $M$ elements in $\{{\bf K}_1\}$.
Since not all of these $M$ elements are in zone A,  we can make each of them inside zone A by subtracting an appropriate reciprocal vector ${\bf G}_{{\bf K}_1}\in\{{\bf G}\}$.
Thus, we define a set $\left\{ {\bf k}_1|{\bf k}_1={\tilde {\bf K}}_1\equiv{\bf K}_1-{\bf G}_{{\bf K}_1} \right.$ in A
for all ${\bf K}_1\in \{{\bf K}_1\}_{{\bf k}_2} \}_{{\bf k}_2}$.
Obviously $\{{\bf k}_1\}_{{\bf k}_2}$ contains $M$ points inside zone A.

{\em Theorem 1.}

(i) The $M$ elements in $\{{\bf k}_1\}_{{\bf k}_2}$ are the counterpoints of ${\bf k}_2$. They
are $M$ different points in zone A.

(ii) All the $M$ eigenstates with the $M$ wave vectors in $\{{\bf k}_1\}_{{\bf k}_2}$ (associated with the unit cell) correspond
to $M$ eigenstates with one wave vector ${\bf k}_2$ in zone B  (associated with the super cell).

(iii) Each band in the band structure associated with the unit cell will split into
$M$ bands in the band structure associated with the super cell.

{\em Proof.} For any wave vector ${\bf k}$ in the ${\bf k}$ space,
${\bf k}_1$ and ${\bf k}_2$ are the counterpoints in zone A and zone B, respectively.
From Eqs. (4), (7) and (8), one has
\begin{eqnarray}
{\bf k}&=&{\bf k}_2+{\bf G'}={\bf k}_1+{\bf G} , \\
{\bf k}_1&=&{\bf k}_2+{\bf G'}-{\bf G}  \nonumber \\
&=&{\bf k}_2+\bar{\bf G}+{\bf G}_1-{\bf G}  \nonumber \\
&\equiv&{\bf k}_2+\bar{\bf G}+{\bf G}_2  \nonumber \\
&=&{\bf K}_1+{\bf G}_2.
\end{eqnarray}
Since ${\bf k}_1$ is in zone $A$, it follows from the definition that ${\bf k}_1\in{\{\bf k}_1\}_{{\bf k}_2}$
(here ${\bf G}_{{\bf K}_1}=-{\bf G}_2$).
Therefore, for any wave vector which has a counterpoint ${\bf k}_2$ in zone B, its counterpoint in zone A
must belong to ${\{\bf k}_1\}_{{\bf k}_2}$.
On the other hand,
all the elements in ${\{{\bf k}_1\}_{\bf k}}_2$ for all possible ${\bf G'}$ (correspond to all possible $\bar{\bf G}\in\{\bar{\bf G}\}$) in Eq. (10) are all the counterpoints of ${\bf k}_2$.
Therefore, the elements in the set ${\{{\bf k}_1\}_{\bf k}}_2$ are exactly all the counterpoints of ${\bf k}_2$.

Consider two different ${\bar{\bf G}}_1,{\bar{\bf G}}_2\in\{\bar{\bf G}\}$. Correspondingly, we have
${\bf K}_1={\bf k}_2+{\bar{\bf G}}_1$ and ${\bf K}_2={\bf k}_2+{\bar{\bf G}}_2$.
From the definition we have
${\tilde{\bf K}}_1-{\tilde{\bf K}}_2={\bf K}_1-{\bf G}_{{\bf K}_1}-({\bf K}_2-{\bf G}_{{\bf K}_2})=
{\bar{\bf G}}_1-{\bar{\bf G}}_2-({\bf G}_{{\bf K}_1}-{\bf G}_{{\bf K}_2})$.
Since
${\bar{\bf G}}_1-{\bar{\bf G}}_2\notin\{{\bf G}\}$ and ${\bf G}_{{\bf K}_1}-{\bf G}_{{\bf K}_2}\in\{{\bf G}\}$,
we know that ${\bar{\bf G}}_1-{\bar{\bf G}}_2$ and ${\bf G}_{{\bf K}_1}-{\bf G}_{{\bf K}_2}$ are different, i.e.,
${\bar{\bf G}}_1-{\bar{\bf G}}_2-({\bf G}_{{\bf K}_1}-{\bf G}_{{\bf K}_2}) \ne 0$, which immediately gives ${\tilde{\bf K}}_1-{\tilde{\bf K}}_2 \ne 0$. This proves that ${\tilde{\bf K}}_1$ and ${\tilde{\bf K}}_2$ are two different points in zone A.
Therefore, the elements in ${\{{\bf k}_1\}_{\bf k}}_2$ are $M$ different points.

Let ${H_{\bf k}}_1$  be the eigenstate for a wave vector in the set ${\{{\bf k}_1\}_{\bf k}}_2$
associated with the unit cell.
From Eqs. (5) and  (10), one has
\begin{eqnarray}
{H_{\bf k}}_1({\bf r})&=&e^{i{{\bf k}_1} \cdot {\bf r}} \sum\limits_{\bf G} {H_{\bf G} e^{i{\bf G} \cdot {\bf r}} } \nonumber \\
&=&e^{i({{\bf k}_2+\bar{\bf G})+{\bf G}_2} \cdot {\bf r}} \sum\limits_{\bf G} {H_{\bf G} e^{i{\bf G} \cdot {\bf r}} }   \nonumber \\
&=&e^{i{\bf k}_2 \cdot {\bf r}} \sum\limits_{\bf G} {H_{\bf G} e^{i({\bf G}+\bar{\bf G}+{\bf G}_2) \cdot {\bf r}}}  \nonumber \\
&=&e^{i{\bf k}_2 \cdot {\bf r}} \sum\limits_{\bf G'} {H_{\bf G'} e^{i{\bf G'} \cdot {\bf r}} }\equiv
{H'_{\bf k}}_2({\bf r}),
\end{eqnarray}
where ${H'_{\bf k}}_2$ is the same eigenstate (with the same field distribution)
but for the wave vector ${\bf k}_2$ in zone B (associated with the super cell).
Therefore,  all the $M$ eigenstates with the $M$ wave vectors in $\{{\bf k}_1\}_{{\bf k}_2}$ (associated with the unit cell) can be
represented by $M$ eigenstates with one wave vector ${\bf k}_2$ in zone A (associated with the super cell).
The $M$  points of ${\bf k}_1$ on any band in the band structure associated with the unit cell
will be on $M$ bands for one ${\bf k}_2$ value in the band structure associated with the super cell.
Generally speaking, one band in the band structure associated with
the unit cell will split into $M$ bands (which may overlap partially and form degenerated eigenstates)
in the band structure associated with the super cell.  The  theorem is thus proved.

{\em Theorem 2.} If a photonic crystal has some kind of point group symmetry,
the eigenstates at the edge of the first Brillouin zone B will be degenerated
in the band structure associated with the super cell. The degree of the degeneracy depends on
both the determinent $M$ of the transform matrix $N$ and the point group symmetry of the photonic crystal.

{\em Proof.}
For a wave vector ${\bf k}_2$ at the edge of zone B, besides that itself
${\bf k}_1={\bf k}_2+{\bf 0}$ (corresponding to ${\bar{\bf G}}_1={\bf 0}$)
is one of its counterpoints at the edge of the zone B,
it may have other counterpoint
${\bf k}'_1={\bf K}_1-{\bf G}_{{\bf K}_1}$   (with ${\bf K}_1={\bf k}_2+{\bar{\bf G}}_2$)   located at somewhere else at the edge of zone B. For nonzero   $\bar{\bf G}$ only those points at the edge of zone B may have counterpoints still
at the edge of zone B and the counterpoints for those points inside zone B will  be outside
zone B (but still inside A according to the definition for counterpoints; note that zone B is inside zone A).
Wave vectors ${\bf k}_1$ and ${\bf k}'_1$ correspond to the same wave vector ${\bf k}_2$ in zone B associated with the super cell.
Sometimes there exists a symmetric operation $\alpha$ (which can be represented by  a matrix for coordinate transformation; then one has $\alpha^{-1}=\alpha^T$)
and the associated operator $T(\alpha)$ (with $T(\alpha)f({\bf r})=f(\alpha^{-1}{\bf r}))$ for the photonic crystal
such that  $\alpha {\bf k}'_1={\bf k}_1$ and
$T(\alpha)\Theta({\bf r})=\Theta(\alpha^{-1}{\bf r})T(\alpha)=\Theta({\bf r})T(\alpha)$.
Assume that  $H_{{\bf k}_1}$ and $H_{{\bf k'}_1}$ are the eigenstates for these two wave vectors, i.e.,
$\Theta H_{{\bf k}_1}  = \frac{{\omega_{\bf k}}_1 ^2}{c^2 }H_{{\bf k}_1}$ and
$\Theta H_{{\bf k'}_1}  = \frac{{\omega_{\bf k'}}_1 ^2}{c^2 }H_{{\bf k'}_1}$.
Since
\begin{eqnarray}
T(\alpha)H_{{\bf k'}_1}({\bf r})&=&H_{{\bf k'}_1}(\alpha^{-1}{\bf
r})=H_{{\alpha \bf k'}_1} ({\bf r})\nonumber \\
&=&H_{{\bf k}_1} ({\bf r}),
\end{eqnarray}
we have
\begin{eqnarray}
\Theta {H_{\bf k}}_1&=&\Theta T(\alpha){H_{\bf
k'}}_1=T(\alpha)\Theta {H_{\bf k'}}_1 \nonumber
\\
&=&\frac{{\omega_{\bf k'}}_1^2}{c^2} T(\alpha){H_{\bf
k'}}_1=\frac{{\omega_{\bf k'}}_1 ^2}{c^2 }{H_{\bf k}}_1,
\end{eqnarray}
we have ${\omega_{\bf k}}_1={\omega_{\bf k'}}_1$. Therefore,
${H_{\bf k}}_1$ and ${H_{\bf k'}}_1$ are two different eigenstates (for different wave vectors ${\bf k}_1$
and ${\bf k'}_1$) with the same eigenvalue.
In the band structure associated with the super cell, these two eigenstates are located at two
bands but have the same wave vector ${\bf k}_2$ and the same eigenvalue. Thus, they are degenerated states.
Since we do not assume any specific form for $\Theta$ in the above proof, the
theorem is valid in any dimensional space (and for any polarization in the 2D case).

In the next section, we will illustrate these degeneracy theorems with some numerical examples and
use the degeneracy analysis to explain PBGs for some known 2D
photonic crystals and create large band gaps  by breaking
the symmetric properties of the photonic crystal.

\section{Numerical results}
\begin{figure}
\includegraphics[width=3.5in]{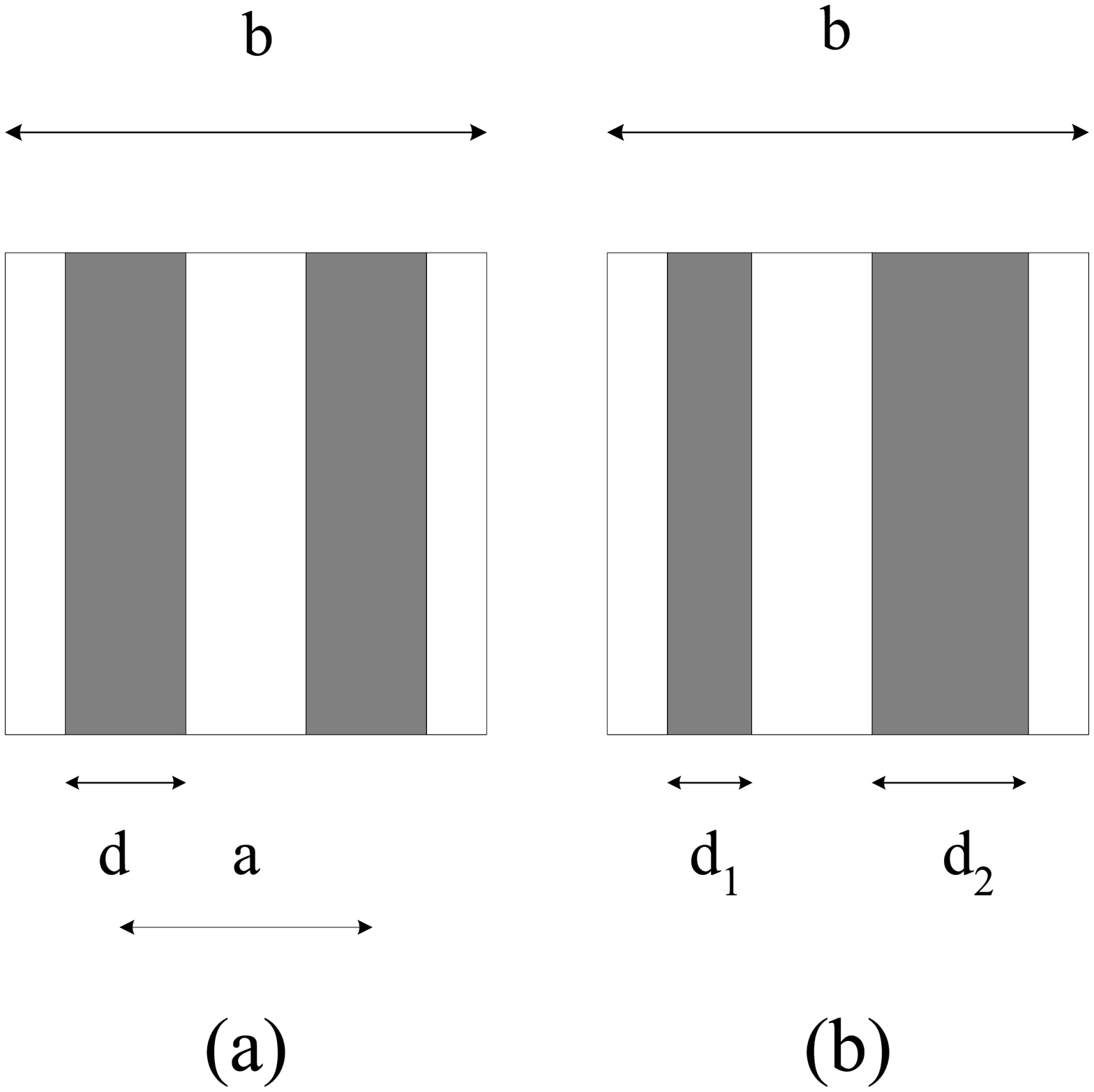}
\caption{\label{fig:figure1}A 1D photonic crystal  consisting of
alternating layers of  two different materials. (a) The super cell
(including 2 unit cells) of the photonic crystal. (b) The symmetry
of  the super cell is broken (by changing the widths of the two
dielectric layers while keeping the position of both unchanged) to
form a unit cell for a new  photonic crystal. }
\end{figure}
First we give a one-dimensional (1D) example.
Fig. 1 is a 1D photonic crystal consisting of alternating layers of materials with two different
dielectric constants ($\varepsilon_1=13  $ and $\varepsilon _2=1 $). We can select a periodic region (a super cell) to include 2 unit cells as shown in Fig. 1(a).
The band structure associated with the unit cell and the band structure associated with  the super cell (with $N=2$)
are given in the same figure (see
Fig. 2(a)), where the frequency and the wave vector are normalized with the same
constant $a=1$ in order to make them comparable. For this case, we have
$\{\bar G\} \equiv \{ {\bar G}_1, {\bar G}_2 \} =\{0,0.5(2\pi /a)\}$ and $M=2$.
\begin{figure}
\includegraphics[width=3.5in]{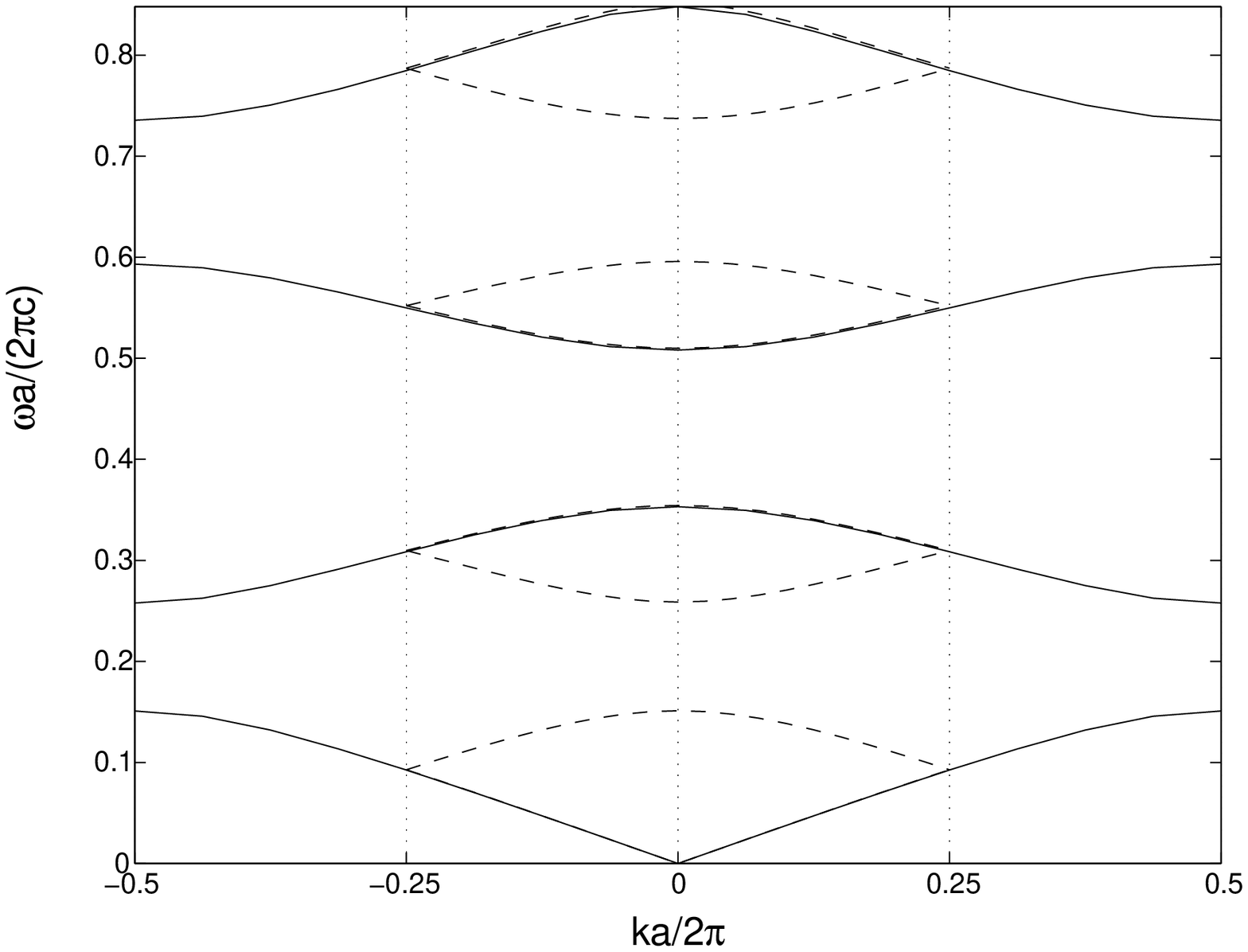}
\includegraphics[width=3.5in]{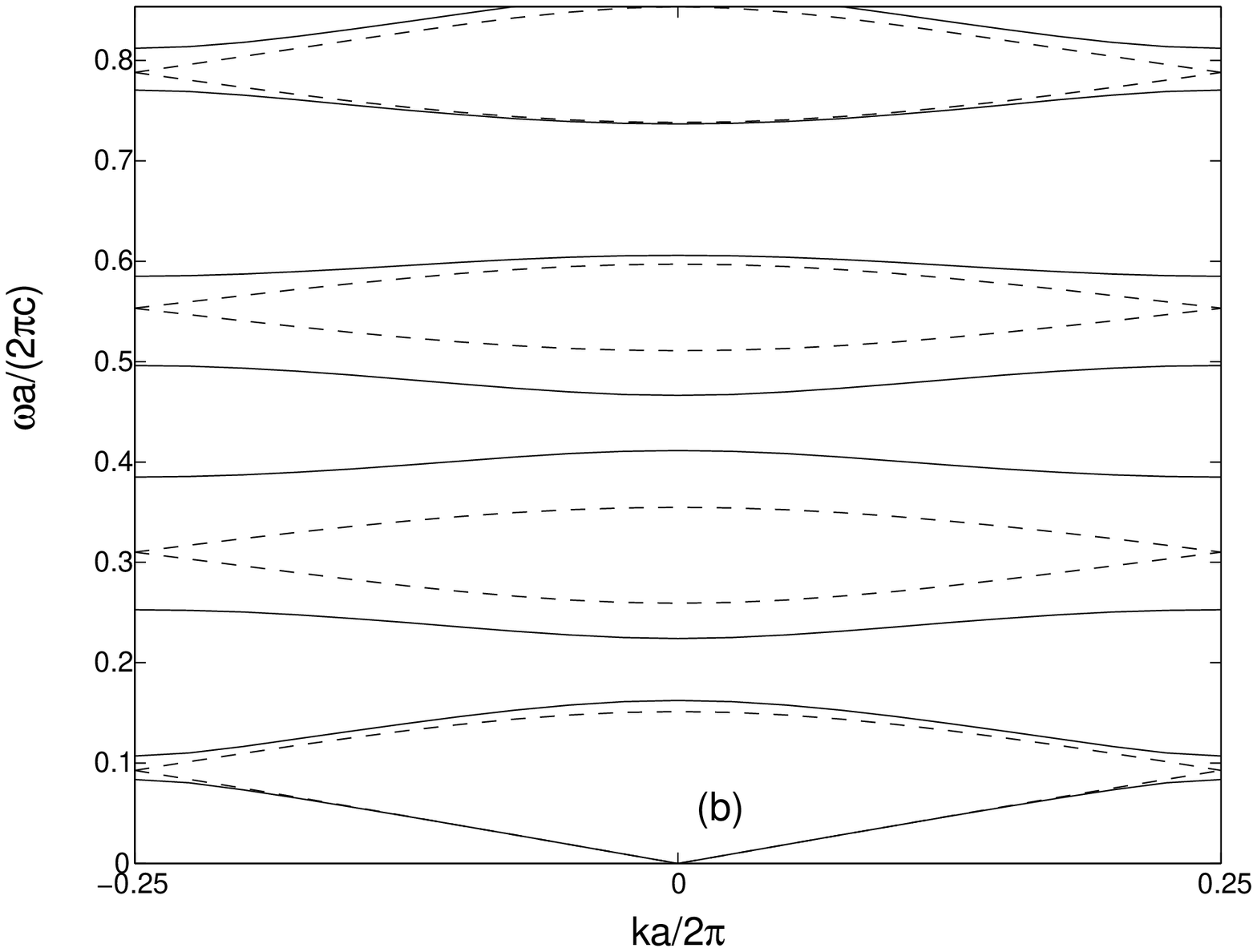}
\caption{\label{fig:figure2}The corresponding band structures of
the 1D photonic crystals with  $\varepsilon_1=13,
\varepsilon_2=1$. (a) The solid lines are for the band structure
associated with the unit cell and the dashed lines give the  band
structure associated with  the super cell.  Here we choose $b=2a$
and $d=0.5a$ for Fig. 1(a). (b) The solid  lines give the band
structure  for a new photonic crystal with a unit cell shown  Fig.
1(b) (here we choose $d_1=0.3a$ and $d_2=0.7a$). The dashed lines
are for the band structure for Fig. 1(a) before the symmetry of
the super cell is broken. }
\end{figure}
From Fig. 2(a) one  sees that the eigenvalues (associated with the original unit cell) for the wave vectors outside the first Brillouin
zone B (associated with the super cell) have their counterpoints  in zone B in the band structure associated with the super cell.
As expected, each band (solid line) associated with the unit cell corresponds to two bands (dashed lines) associated with  the super cell.
Since the center point of the super cell in Fig. 1(a) is mirror symmetric, one has $\alpha=-1$.
At the edge of zone B,  the two wave vectors $k'_1=0.25(2\pi /a)$ and $k_1=-0.25(2\pi /a)=\alpha k'_1$
correspond to the same wave vector $k_2=-0.25(2\pi /a)$ (note that $k_2+ {\bar G}_1=k_1, k_2+ {\bar G}_2=k'_1$).
Thus, these two eigenstates
are  degenerated in the band structure associated with the super cell.
Each eigenstate at $k=\pm 0.25(2\pi /a)$  is formed by two degenerated states  in the band structure associated with the super cell.
If one breaks the point group symmetry with respect to the center point of the super cell
of the photonic crystal by changing the size of the inclusion media,
one obtains a new photonic crystal as shown in Fig. 1(b).
Since the resulted photonic crystal is still mirror symmetric respect to the center point of
the inclusion media, the symmetry breaking with respect to
the center point of the super cell does not change the point group symmetry of
the photonic crystal as a whole.
\begin{figure}
\includegraphics[width=3.5in]{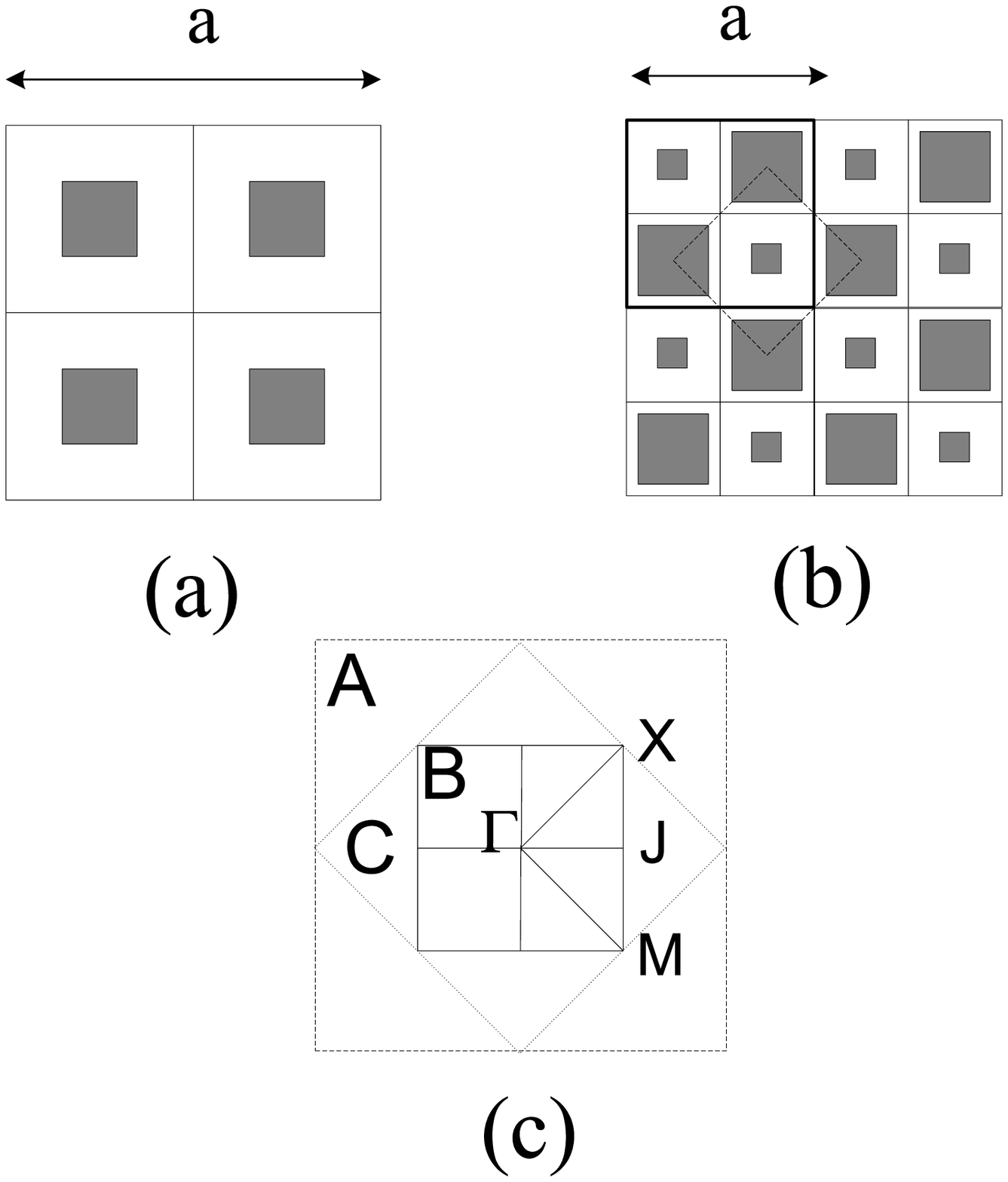}
\caption{\label{fig:figure3}The case for the square lattice of
square dielectric rods. (a) The super cell including 4 unit cells.
(b) The symmetry of the super cell is broken as  two square rods
increase in size and  the other two rods  decrease in  size . The
symmetry-broken super cell is marked by the thick solid lines. (c)
The first Brillouin zone A (marked by the dashed lines) associated
with the original unit cell   and the first Brillouin zone B
(marked by the solid lines)  associated with the super cell of
Fig. 3(a). The first  Brillouin zone C  associated with the new
unit cell (marked by the dashed  lines of  Fig. 3(b))  of the new
photonic crystal of Fig. 3(b) is marked by the  dotted lines.
$\Gamma,X,J,M$ are the symmetry points. } \label{domain}
\end{figure}
However because of the symmetry breaking, the translation group symmetry changes which
leads to a larger unit cell.
The corresponding
transform matrix between the super cell and the new unit cell changes from $N=2$ to $N=1$.
The band structure is shown by the solid lines  in Fig. 2(b), where one sees that the degeneracy disappears
(since $M=1$ for this new photonic crystal and consequently there is only one counterpoint for each wave vector in the Brillouin zone)
and
more band gaps appear.

\begin{figure}
\includegraphics[width=3.5in]{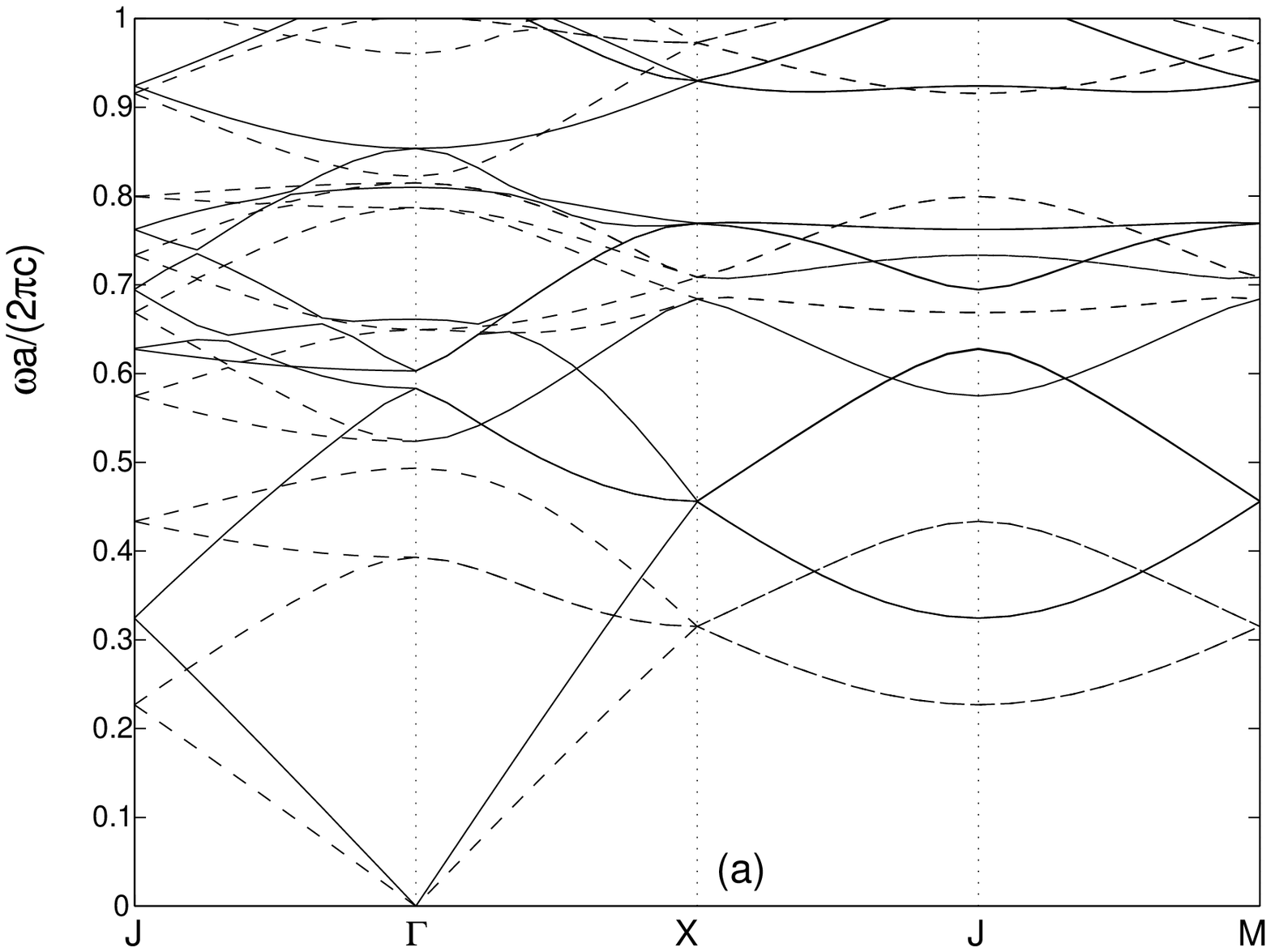}
\includegraphics[width=3.5in]{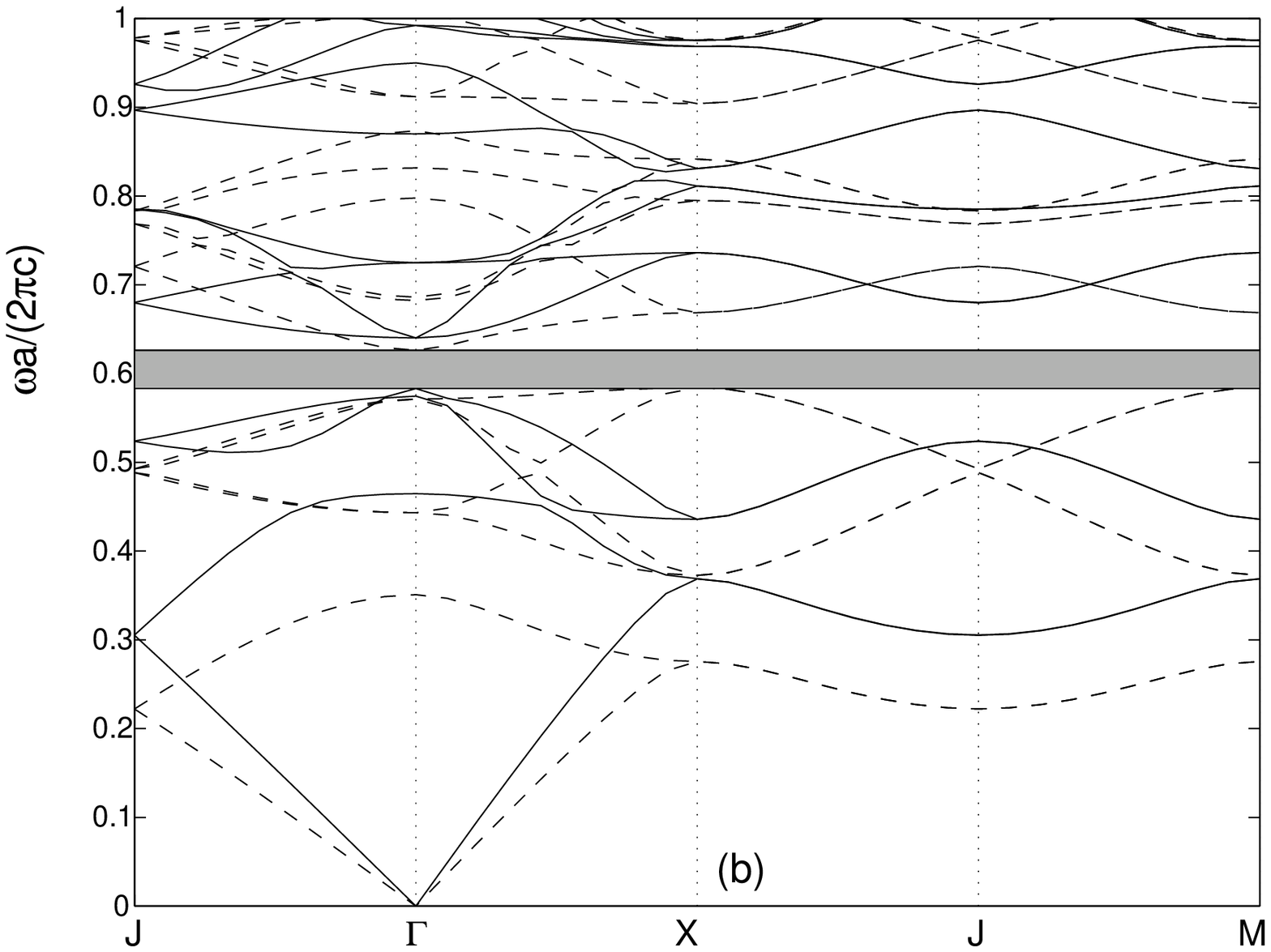}
\includegraphics[width=3.5in]{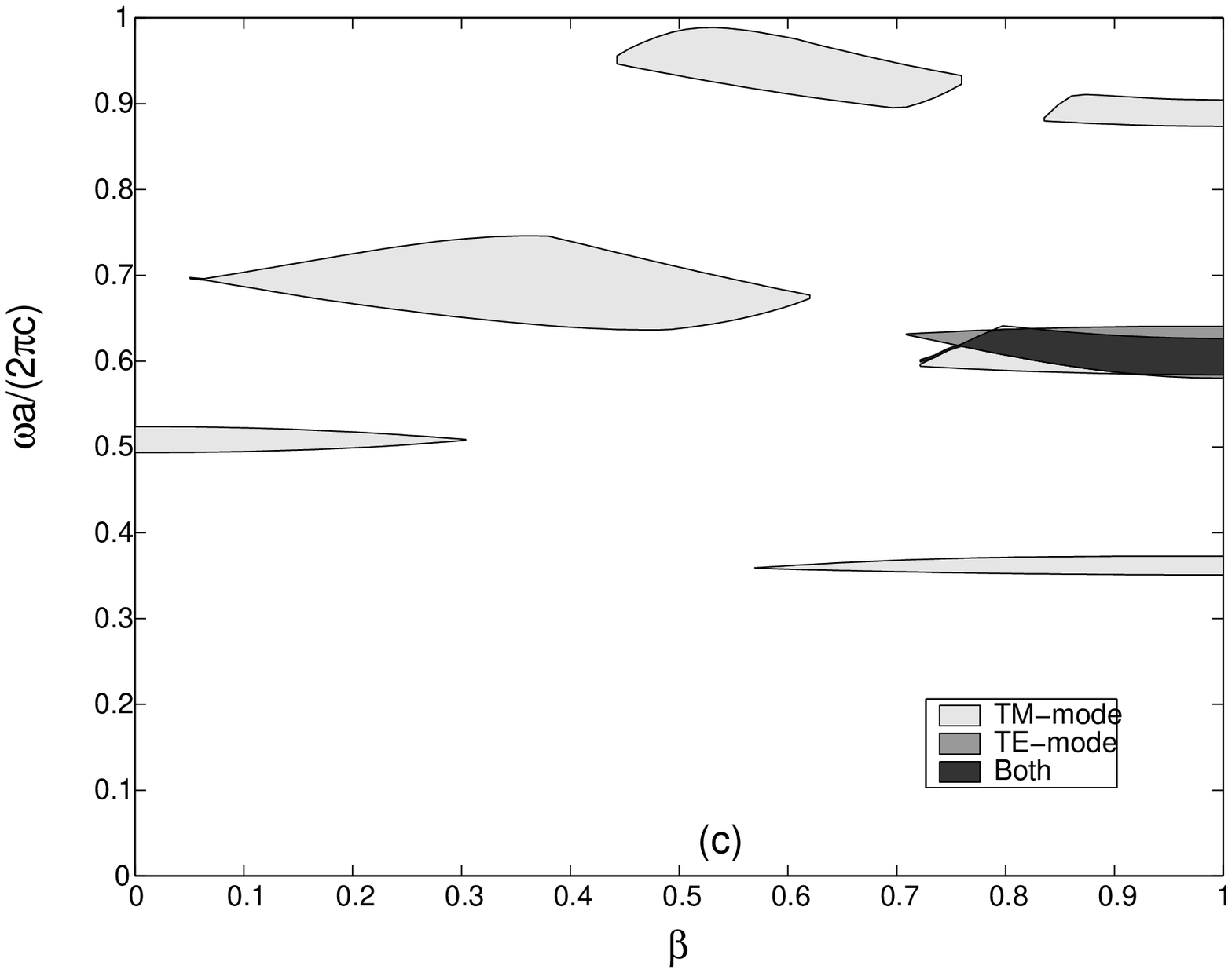}
\caption{\label{fig:figure4}The band structures associated with
Fig. 3 for $f=0.47, \varepsilon=8.9$. The solid lines denote the
H-polarization case and the dashed lines are for the
E-polarization case. They are  calculated by the plane wave
expansion method  (with the inverse rule) with 289 plane waves for
(a) $\beta=0$, i.e., a simple square lattice of square rods; (b)
$\beta=1$, i.e.,  the case of chess board.  The absolute band gap
$\Delta\omega/\omega_c =0.07$ appears at $\omega _c =0.605(2\pi
c/a)$.  (c)The bandgap map for $0\le\beta\le 1$. }
\end{figure}
For the 2D case, if the dielectric inclusions have rectangle shapes,
we can employ the plane wave expansion method with the inverse rule  \cite{inverse} to
calculate the band structure.
It is shown in \cite{inverse,shen} that this method with 225 plane waves can give more accurate
results than a conventional
plane wave
expansion method with even 1681 plane waves. In our calculations, we
use this method with 289 plane waves and the error in the band structure is less than
$0.5\%$.

As an example, we choose $2\times 2$ unit cells as  the super cell. Then
we have
$N^T= \left( {\begin{array}{*{20}c}
   2 & 0  \\
   0 & 2  \\
\end{array}} \right)$, which corresponds to
$\{\bar{\bf G}\}=\{0, {\bf b'}_1, {\bf b'}_2,{\bf b'}_1+{\bf b'}_2\}$ (see the Appendix for a derivation for a general case).
Fig. 3(a) shows a simple square lattice of square dielectric rods.
Alumina is chosen as the dielectric medium and thus $\varepsilon=8.9$.
The filling factor is set to $f=0.47$. Fig. 4(a) gives the band structure associated with the super cell.
Each band (associated with the unit cell) have split into 4 bands in the band structure associated with the super cell.
The point group symmetry is $C_{4v}$.
We use the two mirror symmetries $\alpha_1= \left( {\begin{array}{*{20}c}
   -1 & 0  \\
   0 & 1  \\
\end{array}} \right)$ and
$\alpha_2= \left( {\begin{array}{*{20}c}
   1 & 0  \\
   0 & -1  \\
\end{array}} \right)$  to analyze the degeneracy here.
The eigenstates with wave vectors at the 4 corners (e.g. points $X$ and $M$) of the
Brillouin zone B have  4-fold degeneracy
and the eigenstates with
wave vectors on two opposite edges are of 2-fold degeneracy under these two symmetry operations.
Thus,  in the band structure associated with the super cell one can see that there are
4-fold degeneracies at $X$ and $M$
points
There are also 2-fold accidental degeneracies
for the 4 split bands
(and thus one sees only 3 bands)  in region $\Gamma -X$.
The $E$-polarization and the  $H$-polarization have similar behaviors of degeneracy.

In the super cell, both the size and the position of the inclusions can be changed to break
the point symmetry with respect to the center point of the super cell.
Since the band structure is more sensitive to the inclusion size \cite{Size},
the size of the inclusions will influence significantly the band structure.
In our first example, the  symmetry  is broken
as  two square rods  increase in size and  the other two rods  decrease in  size  in order to
keep the filling
factor $f=0.47$ unchanged.
In the resulted photonic crystal shown in Fig. 3(b), the squares rotate a $45^o$ angle
to form a chessboard structure in the new unit cell
denoted by thickened lines in Fig. 3(b) after the symmetry is broken.
The ratio of the side lengths between the smaller rods to the larger rods is $1-\beta$ with $0\le\beta\le 1$.
When $0<\beta< 1$, it just the case with the smaller square rods being added at the corner of
the simple unit cell.
When $\beta=1$, the side length of the smaller rods is $0$ and only two larger rods exist in the
super cell.
The structure is exactly the chessboard structure reported in \cite{chessboard}.

For $0<\beta\le 1$ shown in Fig. 3(b), since the symmetry with respect to the center point
of each square (which has the highest symmetry)
is still $C_{4v}$, the point group symmetry of the photonic crystal is
still $C_{4v}$.
Similar with 1D case, the symmetry breaking with respect to the center point of the super cell
changes the translation symmetry of the photonic crystal.
The corresponding transform matrix $N$ between the super cell and the unit cell changes from
$N^T= \left( {\begin{array}{*{20}c}
   2 & 0  \\
   0 & 2  \\
\end{array}} \right)$ to
$N^T= \left( {\begin{array}{*{20}c}
   1 & 0  \\
   0 & -1  \\
\end{array}} \right)$
since the unit cell changes to a larger one.
Therefore, the degeneracies of the band structure associated with the super cell will also change.
We take the chessboard ($\beta=1$) as an example to study its band structure (shown in Fig. 4(b)).
From Fig. 4(b) one sees clearly that some degeneracies (including the usual
degeneracies at the edge of zone B and the accidental degeneracies for points at $\Gamma-X$) are
lifted for both the $E$-polarization and the
$H$-polarization. An absolute band gap appears at where the degeneracies are lifted at the edge points of zone B.
To understand this situation, after the unit cell of the photonic crystal in Fig. 3(b)
changes to a larger one (denoted by the dashed line in Fig. 3(b)),
the corresponding first Brillouin of this new photonic crystal is denoted by Zone C in  Fig. 3(c).
The transform matrix between zone C and zone A is
$N^T= \left( {\begin{array}{*{20}c}
   1 & 1  \\
   -1 & 1  \\
\end{array}} \right)$, which corresponds to $M=\det(N^T)=2,  \{\bar{\bf G}\}=\{0, {\bar{\bf b}}'_2\}=\{0, {{\bf b}}'_2\}$
according to the Appendix.
For the point group symmetry of $C_{4v}$,  we can analyze the degeneracy with the two mirror symmetry operators
$\alpha_3=\alpha_1$
and
$\alpha_4=\alpha_2$.
The eigenstates with the 2 wave vectors on the two opposite edges in the Brillouin zone B are 2-fold
degenerate states under these two symmetry operations.
The degeneracy  becomes  only 2-fold  at points $X$ and $M$ now
(as compared to the 4-fold degeneracy in the band structure associated with the super cell)
since $|N^T|=2$ here.
Therefore, degeneracies must disappear at $X$ and $M$ points and each group of 4 bands in  Fig. 4(a) break  at points $X$  and $M$ to form 2 groups with 2 bands in each group
(see Fig. 4(b)).
An absolute band gap $\Delta \omega /\omega _c=0.070$ appears at the mid-frequency (of the band gap)
$\omega = \omega _c  =0.605(2\pi c/a)$.
The reason why they have large absolute band gaps can be explained by the present theory of
super cells.

\begin{figure}
\includegraphics[width=3.5in]{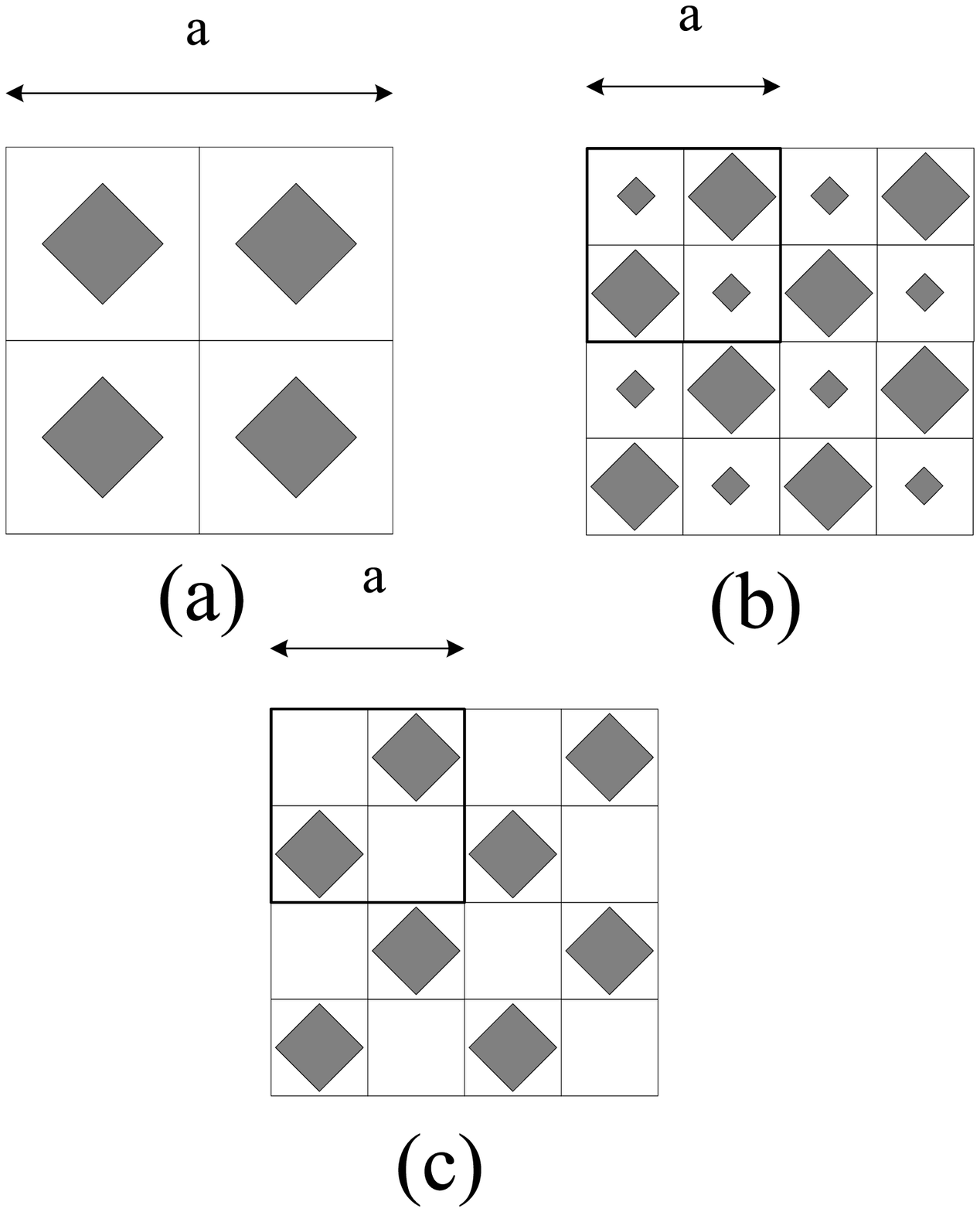}
\caption{\label{fig:figure5}The case of  the chess board
structure. (a) The super cell including 4 unit cells. (b) The
symmetry of the super cell is broken when two square rods increase
in size and the other two square rods decrease in size. (c) The
photonic crystal when the size of the smaller square rods becomes
zero. The structure becomes the simple square lattice of square
rods. }
\end{figure}
Fig. 4(c) is the corresponding gap map when $\beta$ increases from $0$ to $1$.
The situation is similar to the case considered in \cite{Ander} (the only difference is that here we use
square dielectric rods instead of
round air holes). To make a map for the actual procedure of the degeneracy breaking, we take
$\beta$ as the parameter.
The photonic crystal with additional smaller squares ($0<\beta<1$, we call it case 2) is
considered in \cite{Ander} to be the resulted photonic crystal by adding the smaller ones
to corners of the square lattice of the square rods ($\beta=1$, we call it case 3).
Here, we take both case 2 and case 3 to be the results of changing the translation symmetry of
the photonic crystal when the additional
square is of equal sized ($\beta=0$, we call it case 1).
They have the same degeneracy breaking properties with the chessboard structure as mentioned above.
Thus larger absolute band gaps can be expected similarly by choosing an appropriate value of $\beta$.
When $\beta\ge 0.76$, an absolute band gap appears around $\omega=0.6(2\pi c/a)$.
It is more useful to use $\Delta \omega/\omega _c$
to describe the PBGs due to the scaling property of a photonic crystal.
In the band structure associated with the super cell, $\Delta \omega/\omega _c$ remains almost unchanged.
A maximum $\Delta \omega/\omega _c=0.071$ occurs when  $\beta=0.93$.

\begin{figure}
\includegraphics[width=3.5in]{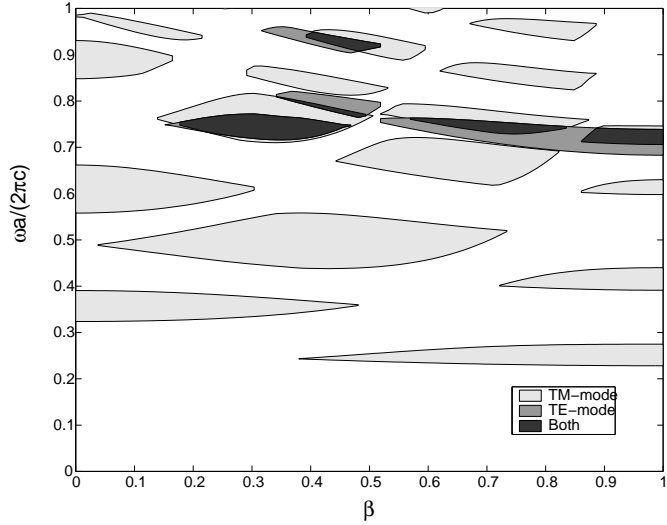}
\caption{\label{fig:figure6}The gap map for Fig. 5 with
$0\le\beta\le 1$ ($1-\beta$ is the side length  ratio of the
smaller squares to the larger ones). }
\end{figure}
Following the same procedure, from a chessboard
photonic crystal shown in Fig. 5(a) we can obtain a photonic crystal formed with square dielectric rods of two different sizes (see Fig. 5(b)) .
The ratio of the side lengths between the smaller rods and  the larger rods is $1-\beta$ with  $0\le\beta\le 1$.
Fig. 6 gives the gap map when the filling ratio is fixed to  $f=0.35$ with $0\le\beta\le 1$ and the inclusion
material has a dielectric constant $\varepsilon=11.4$. From Fig. 6, one sees  that there is no absolute band gap
for the chessboard case (when $\beta=0$) in the frequency range of $0\le\omega\le 2\pi c/a$.  When $\beta =1$, the structure becomes the simple square lattice of square rods of the same size (see Fig. 5(c)), which has an absolute
band gap $\Delta \omega/\omega_c=0.0453$ with the mid-frequency $\omega_c=0.7231(2\pi c/a)$.
It is thus not surprising that with appropriate parameters an absolute photonic band gap exists for the 2D square lattice of
square dielectric rods as considered in \cite{square}.
The  maximal gap of $\Delta\omega/\omega_c=0.0717$ (much larger than the case of inclusions with single size) occurs  at $ \omega  = \omega_c=0.7449(2\pi c/a)$ when the ratio of the size lengths for the two inclusion rods is $\beta=0.31$.

\begin{figure}
\includegraphics[width=3.5in]{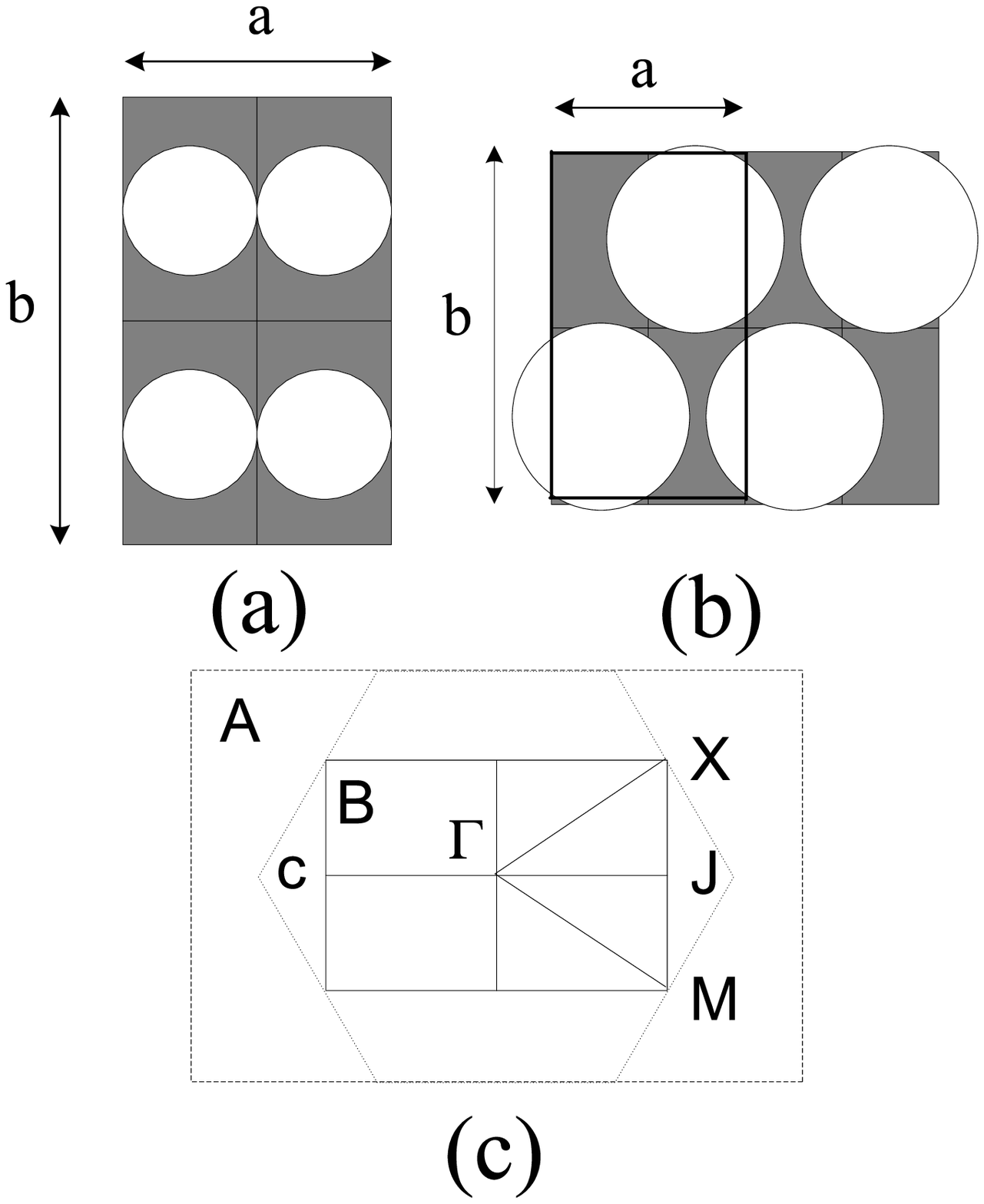}
\caption{\label{fig:figure7}The case for a  rectangular lattice of
air holes. The ratio of the two side lengths is $\sqrt 3$. (a) The
super cell including 4 unit cells. (b) The symmetry of  the super
cell is broken when two diagonal air holes reduce size to zero and
the other two diagonal air holes increase in size. The
symmetry-broken super cell is marked by the thick solid lines. (c)
The first Brillouin zone A (marked by the dashed lines) associated
with the original unit cell   and the first Brillouin zone B
(marked by the solid lines)  associated with the super cell of
Fig. 7(a).  The first Brillouin zone C  associated with the new
unit cell (marked by the dashed  lines of  Fig. 7(b)) of the new
photonic crystal of Fig. 7(b)  is marked by the  dotted lines.
$\Gamma,X,J,M$ are the symmetry points. }
\end{figure}
\begin{figure}
\includegraphics[width=3.5in]{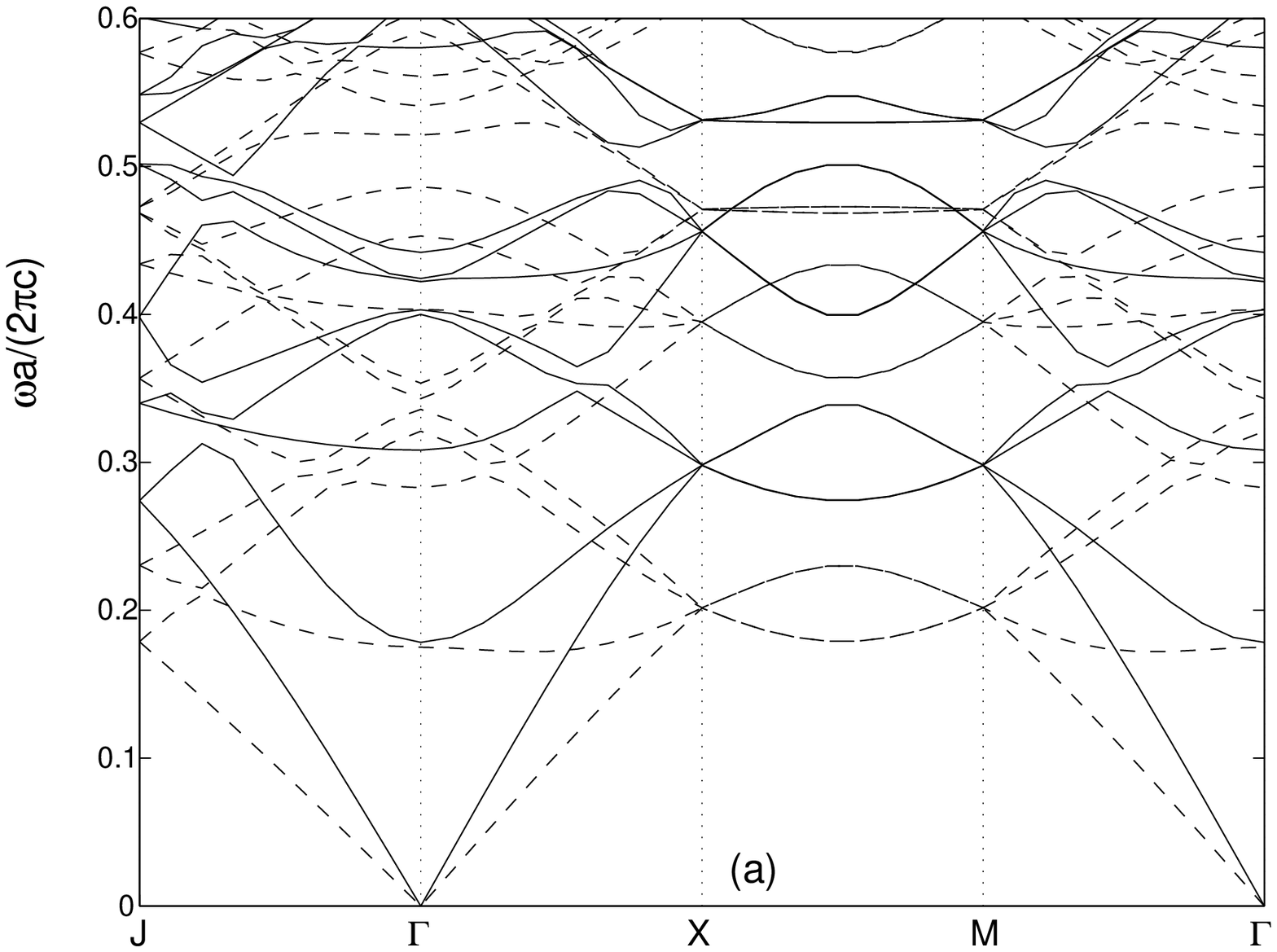}
\includegraphics[width=3.5in]{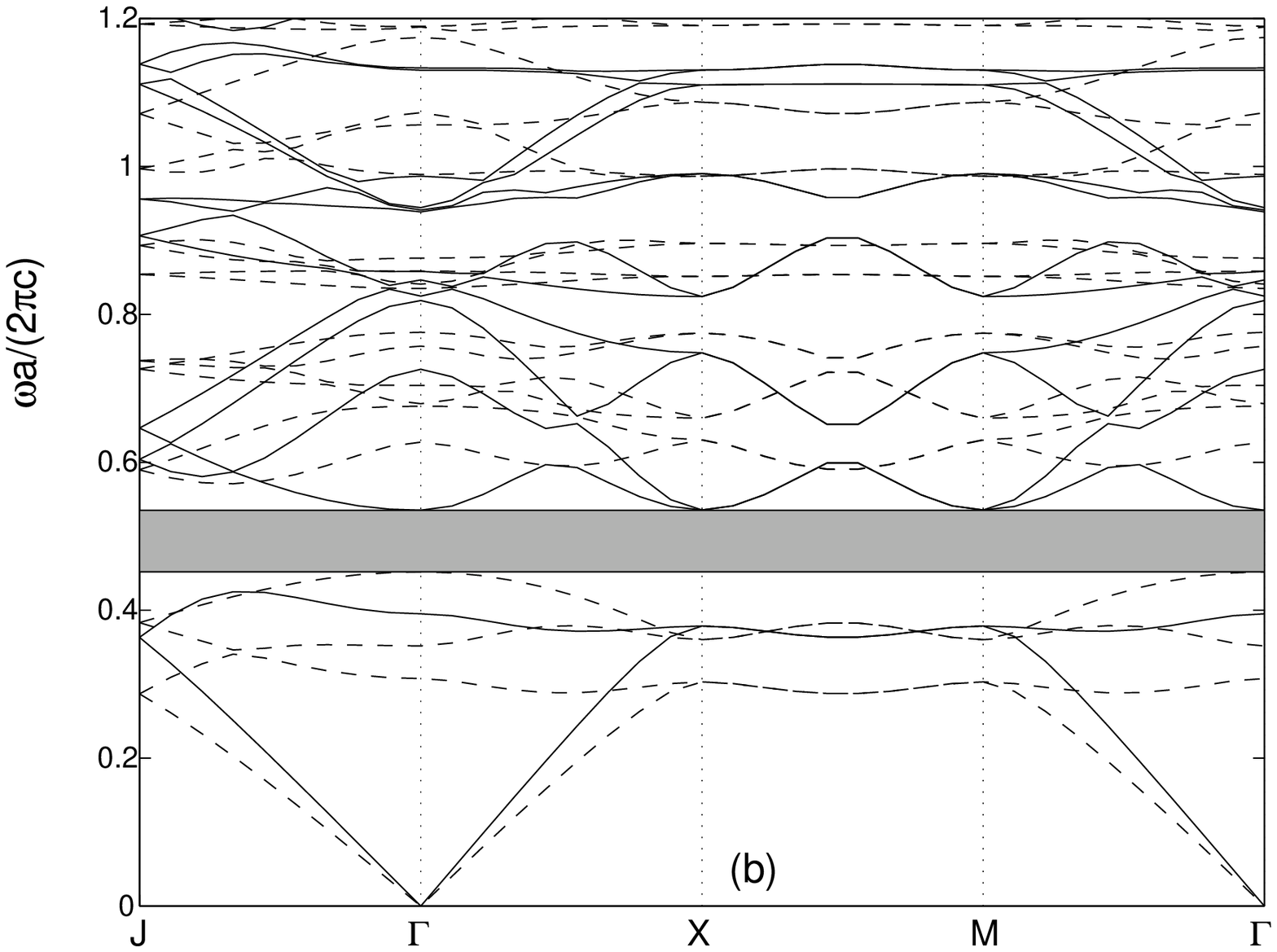}
\caption{\label{fig:figure8}Band structures calculated with the
plane wave expansion method (with 961 plane waves). The relative
permittivity for the background medium is $\varepsilon=13$.  The
solid lines denote the H-polarization case and the dashed lines
are for the E-polarization case. (a) The band structure associated
with the super cell of  Fig. 7(a). The radium of the air holes is
$r=0.5a$. (b) The band structure associated with the
symmetry-broken super cell (marked by the thick solid line in Fig.
7(b)) The filling factor is $f=0.836$. The absolute band gap
$\Delta\omega/\omega _c =0.169$ appears at mid-frequency $\omega
_c =0.4936(2\pi c/a)$. }
\end{figure}
A triangular lattice of air columns (see Fig. 7(b)) has been found to have a large
absolute band gap. The rule of thumb \cite{mold} can be employed to give a reasonable explanation.
Here we explain how a gap appears from the view of changing the translation symmetry.
Although the triangular lattice has a high symmetry, it can be viewed as the result of symmetry breaking from
a super cell of another photonic crystal shown in Fig. 7(a).  From Figs. 8(a) and 8(b) we can see clearly
how the degeneracies are lifted at the edge points $X$ and $M$ (cf. Fig. 7(c)) and an absolute band gap is  created in the band structure associated with the super cell  when the symmetry of  the super cell is broken.

\begin{figure}
\includegraphics[width=3.5in]{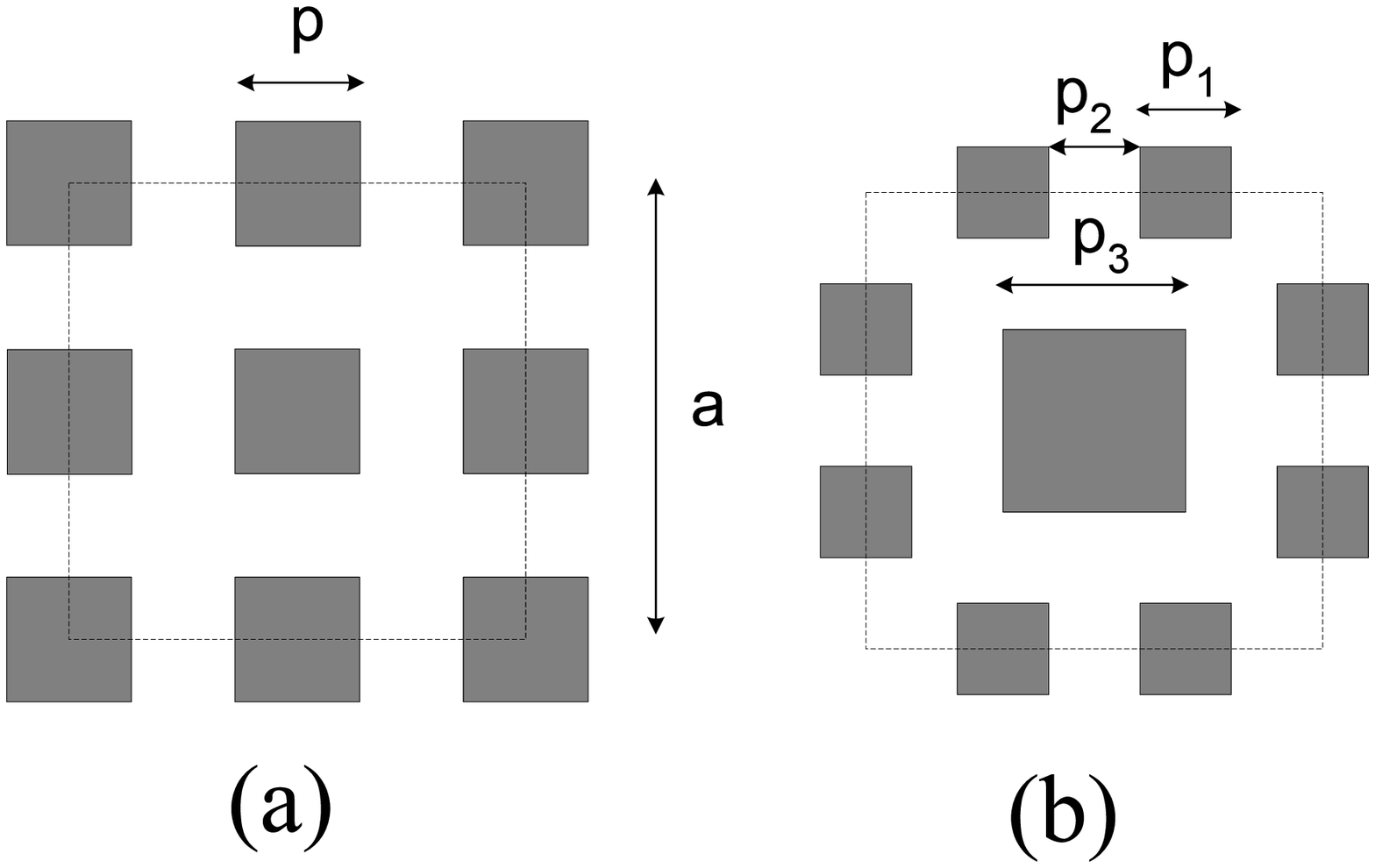}
\caption{\label{fig:figure9}Symmetry breaking of a super cell  by
changing both the sizes and  positions  of the inclusions.   (a)
The super cell including 4 unit cells. (b) The symmetry of  the
super cell is broken when both the sizes (for all the 9 square
rods) and the positions (except the central square rod) of the
square rods are changed (however, with the dielectric filling
factor $ f$ is fixed). }
\end{figure}

\begin{figure}
\includegraphics[width=3.5in]{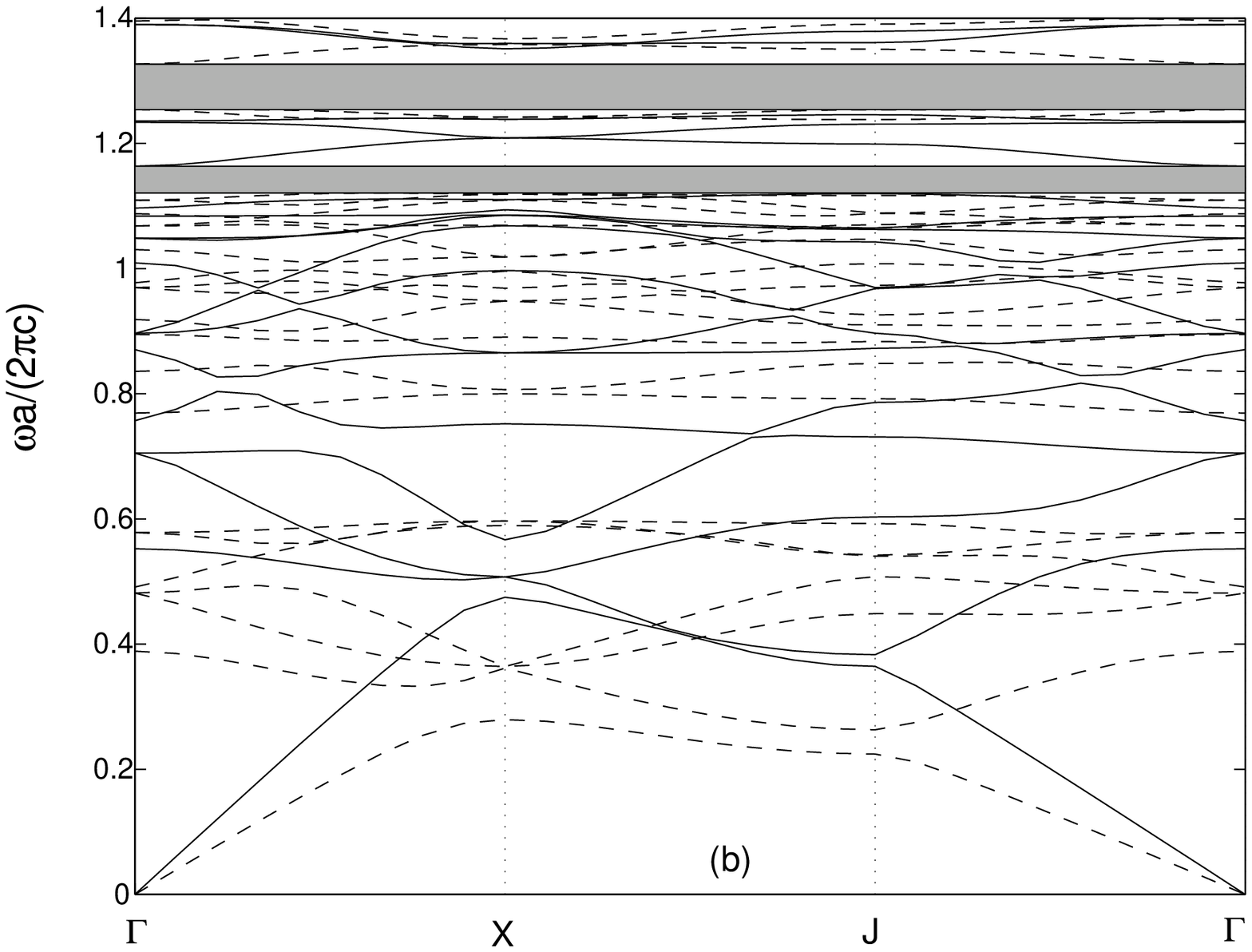}
\caption{\label{fig:figure10}The band structure associated with
the symmetry-broken super cell shown in Fig. 9(b) with $f=0.5132$
and $\varepsilon=11.4$. The solid lines denote the H-polarization
case and the dashed lines are for the E-polarization case. The
other parameters are $p_1=0.194a, p_2=0.236a, p_3=0.374a$. Through
the symmetry-breaking of the super cell,  one obtains a band
structure with two absolute band gaps, namely,  one gap
$\Delta\omega_1=0.0724$ at $\omega_1=1.2905(2\pi c/a)$,  and
another gap $\Delta\omega_2=0.0427$ at $\omega_2=1.1424(2\pi
c/a)$. }
\end{figure}
As a final numerical example, we break the symmetry of a super cell shown in Fig. 9(a) by changing both
the sizes (for all the nine square rods) and the
positions (except the central square rod) of the square rods, however, with the dielectric filling factor $f$ fixed.
Fig. 9(b) is a resulted structure. The band structure with parameters $p_1=0.194a, p_2=0.236a, p_3=0.374, f=0.5132$ are shown in
Fig. 10.  From this figure one sees that the degeneracies are lifted at the edge points and two absolute band gaps are created at higher normalized frequencies, namely,
a large gap $\Delta\omega_1=0.072 (2\pi c/a)$ at
$\omega _c =1.291(2\pi c/a)$ and another gap $\Delta\omega_2=0.043$ at $\omega _c =1.142(2\pi c/a)$.
Note that  it is easier to fabricate a photonic crystal with the absolute band gap
occurring at a higher normalized frequency.

\section{Conclusion}

In the present paper, we have presented a new way of explaining or creating
photonic band gaps through analyzing the  degeneracy
of the band structure associated with a super cell.
The band structure associated with a super cell of a photonic crystal  has
degeneracies at the edge of the first Brillouin zone  if the photonic crystal has some kind
of point group symmetry.
We have analyzed these degeneracies and presented two theorems on degeneracies in the band structure associated with the super cell. These theorems and analysis are valid  in any dimensional space (and for any polarization in the 2D case)
and does not require the investigation of the field distribution.
Photonic band gaps can be created  through lifting these degeneracies by changing the
transform matrix between the super cell and the unit cell which is caused by
changing the translation group symmetry of the photonic crystal.
Many numerical examples have been given in the
present paper to illustrate this.
In the 2D case,   the E-polarization and the H-polarization have the same properties of
degeneracies.
The existence of  photonic band gaps for many known 2D
photonic crystals have been explained through the degeneracy analysis of the band structure
associated with the super cell.
Some photonic crystal structures with large or multiple band gaps have also been  found by breaking
the symmetry of the super cell.

\section*{Acknowledgment}

The partial support of National Natural Science Foundation of China (under a key project grant; grant number  90101024)
and Division of Science and Technologies of Zhejiang provincial government
(under a  key project grant; grant number  ZD0002)  is gratefully acknowledged.

\newpage
\setcounter{equation}{0} \renewcommand{\theequation}{A\arabic{equation}}
\section*{Appendix: The proof of the Lemma and a method to find the set $\{\bar{\bf G}\}$}

We can take 3 kinds of primary transformations for  the integer  matrix $N$ while keeping  the absolute value of $\det(N)$ unchanged.
The first transformation is to multiply a column or row by $\pm 1$.
The second transformation is to interchange two columns or rows.
The third transformation is to add one column or row with $k$ times of
another column or row ($k$ is an integer). Each primary transformation
can be expressed with a left or right multiplication
of the matrix $N$ by an integer matrix.
Furthermore, the determinants of these integer matrices (associated with these primary transformations) are $1$ and
their inverses are also integer matrices.
Under these primary transformations \cite{book2}, the determinent of the matrix is kept unchanged and
the resulted matrix $N'$ is still an integer matrix.

Here we give a specific procedure for obtaining a diagonal matrix $N'=D$ by taking these transformations.
First, we interchange the column or the
row with the second kind of transformation to make $N_{11}$ the minimal
among $N_{1i}$ and $N_{i1}$ , $i=1,2,3$. Then we make $N_{11}$ positive
(if it is negative) with the first kind of transformation. If all the
integers $N'_{1i}$ and $N'_{i1}$ ($i=1,2,3$) are divisible by $N'_{11}$, we can make all $N'_{1i}$ and $ N'_{i1}$ (except
$N'_{11}$) zero by using
the third kind of transformation. If any of $N'_{1i}$ or $N'_{i1}$
is not divisible by $N'_{11}$, its column or row can be subtracted with $k$ times of $N_{11}$
so that the remaining component at $N'_{1i}$ or $N'_{i1}$ position becomes
smaller than $N'_{11}$. The column or row is then interchanged with the first column or the first row
(with $N_{11}$) to make
$N'_{11}$ smaller. This can be done repeatedly until all
$N'_{1i}$ and  $N'_{i1}$ are divisible by $N'_{11}$ or $N'_{11}=1$.
Then we can make all $N'_{1i}$ and $N'_{i1}$ but $N'_{11}$ to
be zero. Applying a similar way to $N'_{22}$, we can make $N'_{23}=N'_{32}=0$. Therefore, the integer matrix $ N^T $ is diagonalized as
\begin{eqnarray}
N^T=P^{-1} D Q, \\
D = \left( {\begin{array}{*{20}c}
   {d_1 } & 0 & 0  \nonumber \\
   0 & {d_2 } & 0  \nonumber \\
   0 & 0 & {d_3 }  \nonumber \\
\end{array}} \right), \end{eqnarray}
where $D, P^{-1}, Q$ are all integer matrices. Here $P^{-1}$ is the
inverse of $P$ and det$(P^{-1})= det(Q)=1$. Thus, we have
$\det(D) = d_1d_2d_3= \det( N^T )= M$.

From Eq. (5) we have
\begin{eqnarray}
{\bf b}_i=\sum\limits_{j = 1}^3{N^T_{ij} {\bf b}'_j}=\sum\limits_{j = 1}^3{(P^{-1}DQ)_{ij} {\bf b}'_j}.
\end{eqnarray}
We can define  two other basic vectors $\bar{\bf b}$ and ${\bar{\bf b}}'$
by $\bar{\bf b}_i=\sum\limits_{j = 1}^3 P_{ij}{\bf b}_j$ and ${\bar{\bf b}}'_i=\sum\limits_{j = 1}^3 Q_{ij}{\bf b'}_j$. Below we will find $\{\bar{\bf G}\}$ in terms of the vectors ${\bar{\bf b}}'_i$.
It follows from Eq. (14) that
\begin{eqnarray}
\bar{\bf b}_i=\sum\limits_{j = 1}^3{D_{ij} {\bar{\bf b}}'_j}=\sum\limits_{j = 1}^3{\delta_{ij}d_j {\bar{\bf b}}'_j}.
\end{eqnarray}Since
\begin{eqnarray}
& &
\sum\limits_{i = 1}^3 n_i{\bf b}_i=\sum\limits_{i,j = 1}^3n_i{P^{-1}}_{ij}{\bar{\bf b}}_j=
\sum\limits_{j = 1}^3 n'_j{\bar{\bf b}}_j\in\{\sum\limits_{i = 1}^3 n_i{\bar{\bf b}}_i\}, \\
& & \sum\limits_{i = 1}^3 n_i{\bar{\bf b}}_i=\sum\limits_{i,j = 1}^3 n_iP_{ij}{\bf b}_j=
\sum\limits_{j = 1}^3 n'_j{\bf b}_j\in\{\sum\limits_{i = 1}^3 n_i{\bf b}_i\},
\end{eqnarray}
we see that the set $\{\sum\limits_{i = 1}^3 n_i{\bar{\bf b}}_i\}=\{\sum\limits_{i = 1}^3 n_i{\bf b}_i\}$ ($n_i,n'_i$ are
arbitrary integers).
Similarly, we can show that
\begin{eqnarray}
\{\sum\limits_{i = 1}^3 n_i{\bar{\bf b'}}_i\}=\{\sum\limits_{i = 1}^3 n_i{\bf b'}_i\}.
\end{eqnarray}

Therefore, the sets ${\{\bf G\}}$ and ${\{\bf G'\}}$ can be written as$\{{\bf G}|{\bf G}=\sum\limits_{i = 1}^3n_i\bar{\bf b}_i=\sum\limits_{j = 1}^3n_id_i{\bar{\bf b}}'_i\}$ and
$\{{\bf G'}|{\bf G'}=\sum\limits_{i = 1}^3n_i{\bar{\bf b}}'_i\}$.

Since any integer $n_i$ can be written as $n_i=l_id_i+m_I$ with $0\le m_i\le d_i-1$, we see that the set $\{\bar{\bf G} | \bar{\bf G}=\sum\limits_{i = 1}^3 {m_i{\bar{\bf b}}'_i},
m_i=0,...,d_i-1\}$.
Obviously, we have $\{\bar{\bf G}\}\cap\{{\bf G}\}={\bf 0}$ and
there are $d_1d_2d_3=M$ possible combinations of $(m_1,m_2,m_3)$ for $\bar{\bf G}$ (i.e., the set $\{\bar{\bf G}\}$ has $M$ elements).
Also, for any two of them, e.g. ${\bar{\bf G}}_1=\sum\limits_{i=1}^3{m_i{\bar{\bf b}}'_i}$,
$\bar{\bf G}_2=\sum\limits_{i=1}^3{m'_i{\bar{\bf b}}'_i}, 0\le m_i\le d_i-1, 0\le m'_i\le d_i-1$,
the difference ${\bar{\bf G}}_1-{\bar{\bf G}}_2=\sum\limits_{i=1}^3{m_i{\bar{\bf b}}'_i}-
\sum\limits_{i=1}^3{m'_i{\bar{\bf b}}'_i}=\sum\limits_{i=1}^3{(m_i-m'_i){\bar{\bf b}}'_i}$
does not belong to $\{{\bf G}\}$ since $0\le | m_i-m'_i | \le d_i-1$.

Obviously, any ${\bf G'}\in\{{\bf G'}\}$ can be written as
\begin{eqnarray}
{\bf G'}&=&\sum\limits_{i = 1}^3n_i{\bar{\bf b}}'_i=\sum\limits_{i = 1}^3 {(l_id_i{\bar{\bf b}}'_i+m_i{\bar{\bf b}}'_i)}, \nonumber \\
&=&{\bf G}+\bar{\bf G},
\end{eqnarray}
where ${\bf G}\in\{{\bf G}\}$ and $ {\bf \bar G}\in\{\bar{\bf G}\}$.

As a numerical example in the 2D case, we can consider the super cell indicated by the thicken line in Fig. 3(b) where the unit cell is indicated by the dashed line (the super cell includes $2$ unit cells). Obviously, we have ${\bf a'}_1={\bf a}_1+{\bf a}_2$ and
${\bf a'}_2={\bf a}_1-{\bf a}_2$. Then we have
$N^T=\left( {\begin{array}{*{20}c}
   1 & 1  \\
   -1 & 1  \\
\end{array}} \right)$.
Following the above procedure, we obtain
$N^T = \left( {\begin{array}{*{20}c}
   1 & 1  \\
   { - 1} & 1  \\
\end{array}} \right) = \left( {\begin{array}{*{20}c}
   1 & 0  \\
   1 & 1  \\
\end{array}} \right)\left( {\begin{array}{*{20}c}
   1 & 0  \\
   0 & 2  \\
\end{array}} \right)\left( {\begin{array}{*{20}c}
   1 & { - 1}  \\
   0 & 1  \\
\end{array}} \right),$
$D= \left( {\begin{array}{*{20}c}
   1 & 0  \\
   0 & 2  \\
\end{array}} \right),$
$P^{ - 1}  = \left( {\begin{array}{*{20}c}
   1 & 0  \\
   1 & 1  \\
\end{array}} \right),$
$Q= \left( {\begin{array}{*{20}c}
   1 & -1  \\
   0 & 1  \\
\end{array}} \right),$
$P= \left( {\begin{array}{*{20}c}
   1 & 0  \\
   -1 & 1  \\
\end{array}} \right),$
$Q^{ - 1}  = \left( {\begin{array}{*{20}c}
   1 & 1  \\
   0 & 1  \\
\end{array}} \right).$
Therefore,  we have $d_1=1, d_2=2$, and
$\{\bar{\bf G}\}=\{0, {\bar{\bf b}}'_2\}=\{0, {{\bf b}}'_2\}$.
Since $\bar{\bf G}_1-\bar{\bf G}_2={\bf b}'_2$ in this special example and we have the following general form for ${\bf G}=n_1{\bf b}_1+n_2{\bf b}_2=n_1({\bf b}'_1+{\bf b}'_2)+ n_2(-{\bf b}'_1+{\bf b}'_2)=(n_1 -n_2) {\bf b}'_1+(n_1 +n_2) {\bf b}'_2$.
$\bar{\bf G}_1-\bar{\bf G}_2\in\{{\bf G}\}$ will end up with
$n_1=n_2=1/2$ which is  contradictory to the requirement that $n_1$ and $n_2$ are integers. Therefore, for this special example we also see that $\bar{\bf G}_1-\bar{\bf G}_2\notin\{{\bf G}\}$.

\end{document}